\documentclass[twocolumn,aps,prl,showpacs,floatfix,,superscriptaddress]{revtex4}
\usepackage[latin9]{inputenc}
\usepackage{graphicx}
\usepackage{esint}

\makeatletter
\@ifundefined{textcolor}{}
{%
 \definecolor{BLACK}{gray}{0}
 \definecolor{WHITE}{gray}{1}
 \definecolor{RED}{rgb}{1,0,0}
 \definecolor{GREEN}{rgb}{0,1,0}
 \definecolor{BLUE}{rgb}{0,0,1}
 \definecolor{CYAN}{cmyk}{1,0,0,0}
 \definecolor{MAGENTA}{cmyk}{0,1,0,0}
 \definecolor{YELLOW}{cmyk}{0,0,1,0}
}

\usepackage{bm}\usepackage{float}\usepackage{afterpage}

\makeatother

\begin{document}

\title{Kaleidoscope of symmetry protected topological phases in one-dimensional periodically modulated lattices}

\author{Huaiming Guo}
\thanks{hmguo@buaa.edu.cn}
\affiliation{Department of Physics, Beihang University, Beijing, 100191, China}

\author{Shu Chen}
\thanks{schen@aphy.iphy.ac.cn}
\affiliation{Beijing National
Laboratory for Condensed Matter Physics, Institute of Physics,
Chinese Academy of Sciences, Beijing 100190, China}
\affiliation{Collaborative Innovation Center of Quantum Matter, Beijing, China}

\begin{abstract}
We identify the existence of various symmetry-protected topological states in one-dimensional superlattices with periodically modulated hopping amplitudes or on-site potentials, which can be characterized by the quantized Berry phase $\pi$ or the emergence of a pair of degenerate boundary states. It is shown that there may exist three types of topological phases, which are protected by the inversion symmetry, the chiral symmetry, and both of them, respectively, depending on the modulations, the odd or even modulation period. The connection between the hopping and potential modulations is also discussed. Furthermore, we demonstrate that the topological phase protected by the inversion symmetry can be realized in the interacting boson systems trapped in the same superlattices. The results are very possibly studied experimentally in the superlattice systems engineered with state-of-art technologies.

%The experimental realization of the identified symmetry protected topological phase can use fermions or bosons loaded into the optical superlattice with the %modulations in the hopping amplitudes or the on-site potentials, which make it very feasible using the state-of-art techniques.
\end{abstract}


\maketitle
%%%%%%%%%%%%%%%%%%%
\textit{Introduction.-}
%%%%%%%%%%%%%%%%%%%
The one-dimensional (1D) topological phases have attracted intense recent studies due to the experimental progress in hybrid superconductor-semiconductor wires \cite{MF1,MF2}, photonic crystals \cite{op1,op2} and cold atomic gases \cite{cold1}. The classification of 1D free fermion systems covering the time-reversal symmetry, particle-hole symmetry and chiral symmetry has been established and five out of ten symmetry classes are topological \cite{sym1,sym2}. The inversion symmetry has also been included and the classification is considerably modified, such as: the AI class becomes topological; the topological invariant of the BDI class is replaced by a $Z_2$ number \cite{inv1,inv2,inv3}. A generalization of the free fermion result to interacting cases has also been obtained for 1D systems \cite{spt1,spt2,spt3,spt4,spt5,spt6}. Many efforts are devoted to construct models belonging to different classes in order to study different kinds of topological properties \cite{Li,LiuXJ,mod1,mod2,mod3,mod4,mod5}.

While the theory of topological classifications indicates permitted types of topological states, existing topological orders in realistic physical systems are rare. Due to their good tunability, the 1D optical and photonic superlattice systems provide an ideal toolbox for exploring topologically nontrivial states \cite{cold1,op1}. Particularly, recent studies of 1D superlattice and quasiperiodic systems from the topological viewpoint \cite{hap1,op1} have unveiled the relation between these systems and two-dimensional (2D) topological insulators \cite{hap1,AA,op1,hof1,hof2,tknn}, which has  been experimentally confirmed by using optical waveguides \cite{op1}. These studies stimulated the exploration of topological phases in 1D superlattice systems \cite{Goldman,op3,hap2,hap3,hap4,hap5,hap6,hap7,hap8}. Nevertheless, it is worth indicating that most previous studied topological phases in 1D superlattice systems do not belong to standard topological  classification of the tenfold way, i.e., they are generally not symmetry-protected topological (SPT) states as these states are characterized by topological invariants of the 2D parameter space through dimensional extension \cite{hap1,hap4,Qi}.

%The Hofstadter butterfly plays a key role in the theory of the quantum Hall effect, which is the energy spectrum of two-dimensional electrons moving under a %perpendicular magnetic field \cite{hof1,hof2,hof3}. For rational flux $\Phi=p/q$, the spectrum consists of $q$ energy bands, which are gapped except the middle %gap of even $q$. The insulator separated by the $r$-th gap is characterized by the Chern number $c$, which satisfy a Diophantine equation: $r=pc+qs$ with %chosen integer $s$ to let $|c|\leq q/2$. A direct manifestation of nonzero Chern numbers is the existence of edge states in the gap, which usually generate %many crossings \cite{tknn}. Recently it has been shown that the Hafstadter model can be mimicked using 1D systems with a periodical modulation parameter as an %additional dimension \cite{hap1,hap2,hap3,hap4,hap5,hap6,hap7}.
%Then the crossings mentioned earlier find their meanings as a pair of degenerate in-gap boundary states. Since their appearance is usually related to 1D %symmetry topological phases, the enormous crossings suggest a big collection of 1D symmetry protected topological phases.

In this work, we explore SPT states in 1D superlattices with periodically modulated hopping amplitudes or on-site potentials, and identify a series of topological phases protected by the inversion symmetry, the chiral symmetry, and both of them, respectively.
All these SPT states are characterized by the presence of a pair of degenerate in-gap boundary states or nontrivial quantized Berry phase. For the systems with periodically modulated hopping amplitudes, the odd modulations only generate the inverse-symmetry-protected topological states, while all three types of SPT states can exist in the even modulation systems. The systems with periodical potentials only support the topological states protected by the inversion symmetry. It is interesting that topological states protected solely by the inversion symmetry corresponds to the crossings of edge modes in the parameter space of the modulation phase, whereas SPT states with the chiral symmetry exists in a much wider regime of the parameter space.
It is also found that 1D bosonic topological phases protected by the inversion symmetry can be directly realized by loading the interacting bosons into the above lattices. Due to the simplicity of our models and the sophisticated technologies of manipulating optical lattices and photonic crystals,
%the independence on the statistics of the particles in the topological phases,
the 1D superlattice systems are thus expected to be an experimentally accessible platform demonstrating rich SPT phases.

%%%%%%%%%%%%%%%%%%%
\textit{Model with periodically modulated hopping.-} We consider the 1D spinless model with periodically modulated hopping amplitudes described by the following Hamiltonian \cite{su1,su2},
\begin{eqnarray}\label{eq1}
\hat{H}_1=\sum_j (t_{j,j+1} \hat{c}_j^{\dagger}\hat{c}_{j+1}+h.c.),
\end{eqnarray}
where $\hat{c}_j,\hat{c}_j^{\dagger}$ is the fermion annihilation and creation operators on the $j$-th lattice site; $t_{j,j+1}$ is the amplitude of the nearest-neighbor hopping and is periodically modulated, i.e., $t_{j,j+1}=t_{j+T,j+T+1}$ with $T$ the period. For convenience, we also denote $t_{j} \equiv t_{j,j+1}$ for $j=1, \cdots, T $. In the presence of the periodic modulations, the unit cell is enlarged to $T$ and the Brillouin zone is reduced to $[-\frac{\pi}{T},\frac{\pi}{T}]$. Correspondingly, the energy spectrum of the uniform system $E(k)=2t\cos k$ with $t_{j,j+1}=t$ is folded into the reduced Brillouin zone and forms $T$ bands with $T-1$ 1D Dirac points. The inclusion of the modulations usually induces gaps between the adjacent bands and generates insulators at the filling $\frac{n}{T}$ with $n=1,...,T-1$. The insulators exhibit various SPT phases. In the following, we focus on $T=3$ and $4$ and the results can be directly generalized to cases with other periods. We also only consider the systems with the total sizes of an integer multiple of the period $T$. The systems with other sizes can be studied similarly by viewing them as supercells of infinite systems \cite{sm}.
\begin{figure}[htbp]
\centering \includegraphics[width=7.5cm]{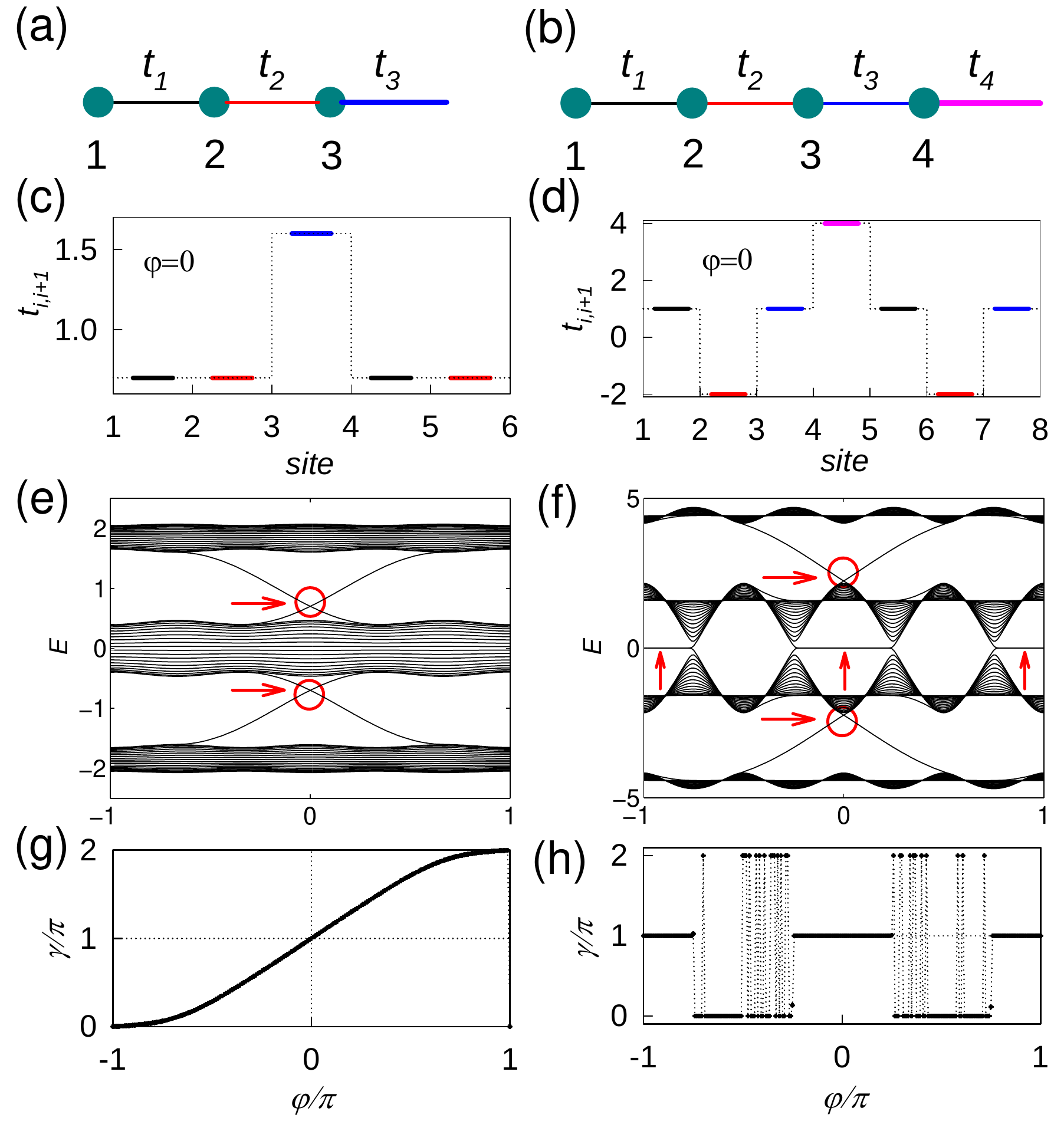} \caption{ (Color online) Schematic illustration of the model Eq.(\ref{eq1}) with the period: (a) $T=3$; (b) T=4. The hopping amplitudes at inversion symmetric point $\varphi=0$: (c) $T=3$; (d) $T=4$. The energy spectrum on an open chain with the length $L=60$: (e) $T=3$; (f) $T=4$. The Berry phase corresponding to the above figure: (g) $T=3$ at $1/T$ filling; (h) $T=4$ at half filling. The parameters $\lambda=0.6$ for $T=3$ cases and $\lambda=3$ for $T=4$ cases are used.}
\label{fig1}
\end{figure}

Firstly we take the cosine modulations $t_{j,j+1}=t[1+ \lambda \cos (\frac{2\pi j}{T}+\varphi)]$ with $t=1$ taken as the energy unit,  which have
period $T$ and phase factor $\varphi$. With this kind of modulations, all bands are gapped at any $\varphi$ for odd $T$, while the system at half filling remains gapless at $\varphi=\frac{2\pi}{T}(n+\frac{1}{2}), n=0,...,T-1$ for even $T$. To give concrete examples, we display schematic structures of superlattices with $T=3$ and $4$ in Fig. \ref{fig1} (a)- (d)  and the corresponding energy spectrum under open boundary conditions in Fig. \ref{fig1} (e) and (f). The continuous edge states traversing the gap at the fillings of $1/T$ and $(T-1)/T$ have a crossing at the inversion symmetric point $\varphi=0$, which indicates
%So 1D topological phase is identified, which manifests itself by the emergence of a pair of degenerate in-gap boundary states.
the emergence of a pair of degenerate in-gap boundary states protected by the inversion symmetry \cite{sm}.
Their appearance is due to the topologically nontrivial property of the 1D bulk system, characterized by the quantized Berry phase $\pi$, which can be calculated via $\gamma=\oint\mathcal{A}(k)dk$
with the Berry connection $\mathcal{A}(k)=i\langle u_{k}|\frac{d}{dk}|u_{k}\rangle$
and $|u_{k}\rangle$ the occupied Bloch states \cite{berry1,berry2}. In Fig. \ref{fig1} (g), we show the Berry phase versus $\varphi$ for the example system with $T=3$ and $\lambda=0.6$. It is clear that the Berry phase is quantized to $\pi$ at $\varphi=0$, serving as a hallmark of the 1D topological state, whereas $\gamma=0$ at the other inversion-symmetry point $\varphi=\pi$ corresponds to a topologically trivial state.
%Regarding the Hamiltonian Eq.(\ref{eq1}) with the cosine modulations, it only has inversion symmetry at $\varphi=0$ and $\pi$, where the Berry phase is %quantized to $\pi$ and $0$, respectively.
Similar topological states protected by the inversion symmetry also exist in superlattice systems with generic $T$ at fillings of $1/T$ and $(T-1)/T$, e.g., as illustrated in Fig. \ref{fig1} (f), states for the system with $T=4$ and $\varphi=0$ at fillings of $1/4$ and $3/4$ are topologically nontrivial.

Generally, the periodic modulation can take any forms, and 1D topological phases may occur in the presence of the inversion symmetries $t_j=\pm t_{T-j}$ with $j=1,...,T-1$. When the inversion symmetry is present, the Berry phase is quantized and can take only the value $0$ and $\pi$ (modulo $2\pi$) \cite{zak,kohn}, with the topological states characterized by the quantized Berry phase $\pi$.
Depending on the values of the hopping amplitudes, the topological phase may be realized at fillings of $1/T$ and $(T-1)/T$. The system with $T=3$ is topological at ${1}/{3}$ and ${2}/{3}$ fillings when $|t_1|=|t_2|$ and $|t_1|<|t_3|$, while the system with $T=4$ is topological at ${1}/{4}$ and ${3}/{4}$ fillings when $|t_1|=|t_3|$ and $|t_2|<|t_4|$.

Another class of 1D topological phases is identified in systems with even $T$s at half filling, which appears in alternating regions separated by the gapless points $\varphi=\frac{2\pi}{T}(n+\frac{1}{2})$. The topological and trivial regions interchange at a critical $\lambda_c$ ($\lambda_c=\sqrt{2}$ for $T=4$) when the gap at half filling closes. The states can be characterized by the Berry phase, which is also quantized to $0$ or $\pi$. The value $\pi$ corresponds to the topological phase, which manifests itself by the existence of a pair of zero boundary modes under open boundary conditions. For general modulations, the 1D topological phase exists in a finite region near $|t_1|=|t_3|$ when $|t_2 t_4|>t_1^2$.

Next we explore the symmetries protecting the above 1D topological phases. The system of $T=4$ is studied and its Hamiltonian in the basis of the momentum space $(\hat{c}_{1,k},\hat{c}_{2,k},\hat{c}_{3,k},\hat{c}_{4,k})^{T}$ writes as \cite{sm}:
\begin{eqnarray}\label{eq2}
\hat{H}_1(k) &=& \frac{t_1+t_3}{2}{I}\otimes \tau_x+\frac{t_1-t_3}{2} s_z\otimes \tau_x \nonumber \\
&+&\frac{t_2+t_4\cos k}{2} s_x\otimes \tau_x+\frac{t_2-t_4\cos k}{2} s_y\otimes \tau_y \nonumber \\
&+&\frac{t_4\sin k}{2} s_y\otimes \tau_x+\frac{t_4\sin k}{2} s_x\otimes \tau_y,
\end{eqnarray}
with $I$ the identity matrix and $s_j, \tau_j (j=x,y,z)$ the Pauli matrices. $\hat{H}_1(k)$ has the chiral symmetries, i.e., $\hat{C}\hat{H}_1(k)\hat{C}^{-1}=-\hat{H}_1(k)$ with the chiral operator $\hat{C}=I \otimes \tau_z$.
For $t_1=t_3$, $\hat{H}_1(k)$ is inversion symmetric, i.e., $\hat{P}\hat{H}_1(k)\hat{P}^{-1}=\hat{H}_1(-k)$ with the inversion operator $\hat{P}=s_x\otimes \tau_x$.
The 1D topological phases at fillings of $1/T$ and $(T-1)/T$ vanish as soon as $\hat{P}$ is broken, as these SPT phases are solely protected by the inversion symmetry. On the other hand, the 1D topological phases at half filling are protected by the chiral symmetry $\hat{C}$. However the inversion symmetric points $\varphi=0$ and $\pi$ are special, where the symmetries $\hat{P}$ and $\hat{C}$ coexist. At the two points the topological phases appear for $\lambda>\lambda_c$. If a next-nearest-neighbor hopping $H_{NNN}=t'\sum_j (\hat{c}_j^{\dagger}\hat{c}_{j+2}+H.c.)$ is added to break the chiral symmetry, the topological phases are broken except at $\varphi=0$ and $\pi$, which implies that the 1D topological phases at $\varphi=0$ and $\pi$ are protected by both the inversion symmetry and the chiral symmetry. Interestingly they can be broken separately, but the topological phase persists. The 1D topological phases protected by the inversion symmetry and both of the two symmetries also exist for $t_1=-t_3$, when the inversion operator becomes $\hat{P'}=s_y \otimes \tau_y$.
%Since the Hamiltonian Eq.(\ref{eq1}) regards spinless fermions and real hopping amplitudes,
We note that the Hamiltonian of Eq.(\ref{eq1}) also has time-reversal symmetry with the time-reversal operator $\cal{T}=\cal{K}$ (the complex conjugate). Thus our results identify some concrete models exhibiting 1D SPT phases belonging to the BDI class, the AI class and the corresponding ones with inversion symmetry. %All these topological phases can be characterized by the Berry phase, or the emergence of a pair of degenerate in-gap boundary states under open boundary %conditions.

\begin{figure}[htbp]
\centering \includegraphics[width=7.5cm]{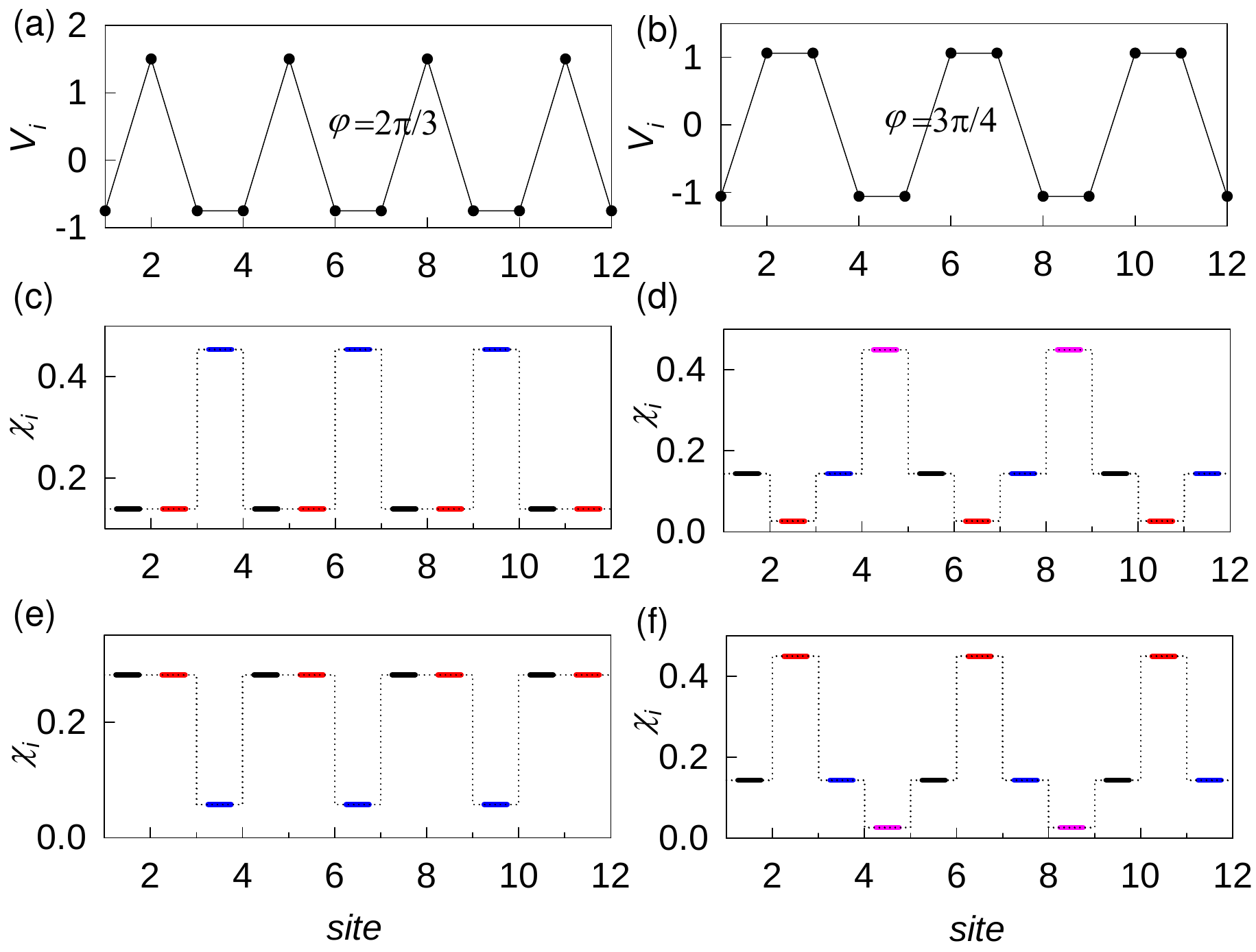} \caption{(Color online) The periodic on-site potential at inversion symmetric point $\varphi=\pi(1-1/T)$  with the period: (a) $T=3$; (b) $T=4$. The effective hopping amplitude $\chi_i$ of the Hamiltonian described by Eq.(\ref{eq3}) for $T=3$ and $\varphi=\frac{2\pi}{3}$ at the filling: (c) $\frac{1}{3}$; (e) $\frac{2}{3}$. $\chi_i$ for $T=4$ and $\varphi=\frac{3\pi}{4}$ at the filling: (d) $\frac{1}{4}$; (f) $\frac{3}{4}$. The parameter $\lambda=1.5$ is used.}
\label{fig2}
\end{figure}

%%%%%%%%%%%%%%%%%%%
\textit{Model with periodically modulated potentials.-}
%%%%%%%%%%%%%%%%%%
We show another kind of spinless models with periodically modulated on-site potentials, which also exhibit many 1D SPT phases. The model is described by the following Hamiltonian,
\begin{eqnarray}\label{eq3}
\hat{H}_2=t \sum_j (\hat{c}_j^{\dagger}\hat{c}_{j+1}+h.c.)+\sum_{j}V_{j} \hat{n}_j,
\end{eqnarray}
where $\hat{n}_j=\hat{c}_j^{\dagger}\hat{c}_j$ is the number operator, and $V_j$ is the strength of the periodically modulated on-site potentials, which satisfies $V_j=V_{j+T}$ with the period $T$. Firstly we consider the modulation with the simple cosine form $V_{j}= \lambda \cos(\frac{2\pi j}{T}+\varphi)$. The 1D topological phase protected by the inversion symmetry is identified at two inversion symmetric points $\varphi=\pi(1-1/T)$ and $\pi(2-1/T)$ for fillings of $1/T$ and $(T-1)/T$. Our calculations show that the Berry phase is quantized to $0$ or $\pi$ at inversion symmetric points for the above fillings and a pair of degenerate states appear in the gap for the open boundary system when the Berry phase of the bulk system is $\pi$. Also the periodic potential can take general forms with inversion symmetry, i.e., $V_j=V_{T+1-j}$ with $j=1,...,T$. Then the 1D topological phase may be realized depending on the values of the potentials. The system with $T=3$ is inversion symmetric for $V_1=V_3$, and is topological for $V_1<V_2$ at $1/3$ filling, whereas $V_1>V_2$ at $2/3$ filling. The system with $T=4$ is inversion symmetric for $V_1=V_4$ and $V_2=V_3$, and is topological for $V_1<V_2$ at $1/4$ filling, whereas $V_1>V_2$ at $3/4$ filling.

The Hamiltonian Eq.(\ref{eq3}) with $T=4$ in the momentum space writes as,
\begin{eqnarray}\label{eq4}
\hat{H}_2(k)&=&t\frac{1+\cos k}{2} s_x\otimes \tau_x +t\frac{1-\cos k}{2} s_y\otimes \tau_y \nonumber \\ \nonumber
&+&t\frac{\sin k}{2} s_y\otimes \tau_x+t\frac{\sin k}{2} s_x\otimes \tau_y +t{I}\otimes \tau_x \\
&+&\tilde{V_2} I \otimes \tau_z+\tilde{V_3} s_z \otimes I+\tilde{V_4} s_z \otimes \tau_z + \tilde{V_1},
\end{eqnarray}
with $\tilde{V_1}={(V_1+V_2+V_3+V_4)}/{4}$, $\tilde{V_2}={(V_1-V_2+V_3-V_4)}/{4}$, $\tilde{V_3}={(V_1+V_2-V_3-V_4)}/{4}$ and $\tilde{V_4}={(V_1-V_2-V_3+V_4)}/{4}$.
$\hat{H}_2(k)$ is inversion symmetric for $V_1=V_4$ and $V_2=V_3$  with the inversion operator given by $\hat{P}=s_x\otimes \tau_x$. Thus the above identified 1D topological phases are protected by the inversion symmetry. Since the Hamiltonian Eq.(\ref{eq4}) has no chiral symmetry, no topological phases appear at half filling.

As both kinds of models display rich SPT phases, we show that the model with periodically modulated potentials can be also understood in terms of the one with periodic hopping amplitudes. Since the periodic potential leads to the distribution of particles varying periodically, the effective hopping amplitudes are affected. To see it clearly, we calculate the effective nearest-neighbor hopping amplitudes defined as $\chi_i=|\langle \hat{c}_i^{\dagger}\hat{c}_{i+1}\rangle|$ ($\langle ...\rangle$ indicates the expectation value in the ground state) for systems of $T=3$ and $T=4$ with $\lambda=1.5$. As shown in Fig.\ref{fig2}, the effective hopping amplitudes are periodically modulated with the period $T$. For $T=3$ and $\varphi=\frac{2\pi}{3}$, we have $\chi_1=\chi_2$ and $\chi_1<\chi_3$ at ${1}/{3}$ filling, which corresponds to a topological phase; whereas at ${2}/{3}$ filling, $\chi_1=\chi_2$ and $\chi_1>\chi_3$, which corresponds to a trivial phase. For $T=4$ and $\varphi=\frac{3\pi}{4}$, we have $\chi_1=\chi_3$ and $\chi_2<\chi_4$ at ${1}/{4}$ filling, which corresponds to a topological phase; on the other hand, we get $\chi_1=\chi_3$ and $\chi_2>\chi_4$ at ${3}/{4}$ filling, which corresponds to a trivial phase. Thus the topological phase induced by periodic potentials can be well understood in terms of the effective hopping amplitudes.

%%%%%%%%%%%%%%%%%%%
\textit{Bosonic topological phases protected by the inversion symmetry.-}
%%%%%%%%%%%%%%%%%%%
\begin{figure}[htbp]
\centering \includegraphics[width=8.5cm]{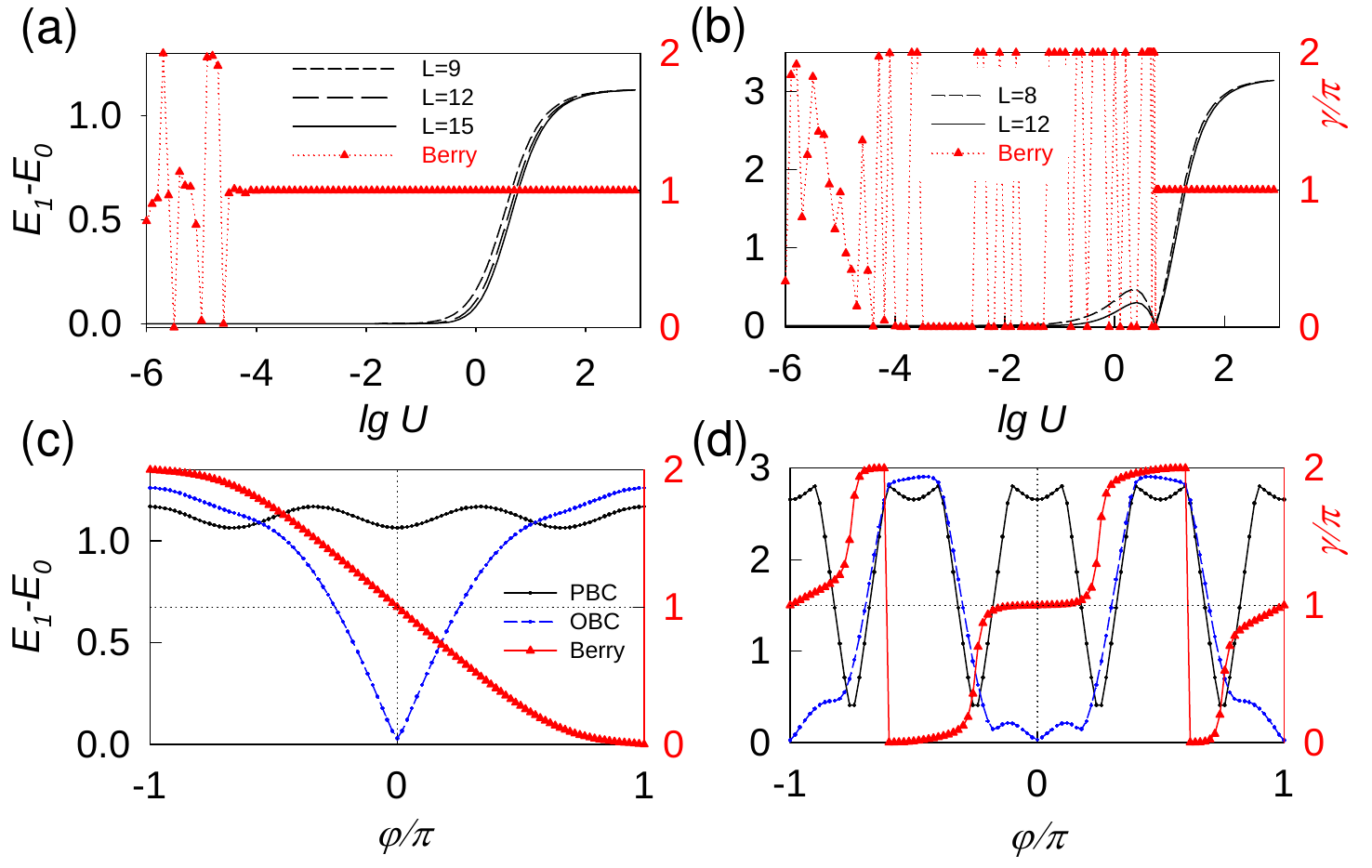} \caption{ (Color online) The energy gap $E_1-E_0$ and the Berry phase $\gamma$ vs $U$ with $\varphi=0$: (a) $T=3$ at $\frac{1}{3}$ filling; (b) $T=4$ at half filling. $E_1-E_0$ and $\gamma$ vs $\varphi$ with fixed $U=50$: (c) $T=3$ at $\frac{1}{3}$ filling; (d) $T=4$ at half filling. The $L=15 (L=12)$ sites are used to calculate $\gamma$ in (a) [(b)] and the quantities in (c) [(d)]. }
\label{fig3}
\end{figure}
We study the topological property of the interacting bosons loaded into the optical superlattice, which is described by a Bose-Hubbard model with periodical modulations of hopping amplitudes \cite{notep}:
\begin{eqnarray}\label{eq5}
\hat{H}_3=\sum_i (t_{i,i+1} \hat{b}_i^{\dagger}\hat{b}_{i+1}+h.c.)
+U\sum_j \hat{n}_j(\hat{n}_j-1)/2,
\end{eqnarray}
where $\hat{b}_j$ ($\hat{b}_j^{\dagger}$) is the bosonic annihilation (creation) operator, $\hat{n}_j=\hat{b}_j^{\dagger}\hat{b}_j$ the number operator of bosons and $U$ represents the strength of on-site interactions. The inclusion of the interaction generally breaks the chiral symmetry.

For interacting bosons trapped in optical superlattices, increasing the repulsive interaction $U$ shall drive the system at commensurate fillings into a Mott insulator. As the on-site interaction does not break the inversion symmetry, one would expect that the bosonic Mott insulator is a topological Mott insulator protected by the inversion symmetry. To see it clearly, we perform exact diagonalization of the Hamiltonian Eq.(\ref{eq5}) with the same parameters used in Fig.\ref{fig1} \cite{exact}.  The energy gaps between the ground state and the first excited state as well as the Berry phase of the system with $\varphi=0$ are calculated as a function of $U$ \cite{note}. As shown in Fig.\ref{fig3}(a) and (b), the gaps begin to develop at small interactions and the resulting Mott insulator is characterized by a nontrivial Berry phase $\gamma=\pi$, where the Berry phase for a many-body system is defined using the twisted boundary phase $\theta$ \cite{hap5,hap6,hap7,twist1}. Though the gaps are generated at other $\varphi$s, the Mott insulators are topological only at the inversion symmetric point $\varphi=0$ for the system of $T=3$ and at $\varphi=0$ and $\pi$ for $T=4$. Apart from these inversion symmetric points, the Berry phase is no longer quantized to $\pi$ as shown in Fig.\ref{fig3}(c) and (d), which clearly unveils the resulting topological Mott phase is protected by the inversion symmetry.

According to the bulk-edge correspondence for topological systems, twofold degenerate edge states are expected to appear on an open chain. As displayed in Fig.\ref{fig3}(c) and (d), the excitation gap gets its minimum at $\varphi=0$ (also $\varphi=\pi$ for the system of $T=4$), which approaches zero in the thermodynamic limit. Here the emergent degeneracy at the inversion symmetric point is related to the boundary excitations of quasiparticles. Particularly, in the hardcore limit $U=\infty$, the system can be mapped to the free fermion model Eq.(\ref{eq1}) via the Jordan-Wigner transformation \cite{jordan}, thus all the three-type SPT phases can all be realized in this limit, which is experimentally accessible by loading the interacting bosons into the corresponding optical superlattices and adiabatically tuning the interaction strength to the strongly interacting limit with the help of Feshbach resonance techniques.
%So the 1D hardcore bosonic topological phases protected by the inversion symmetry can be adiabatically deformed to the Mott insulators at finite $U$, resulting %in the 1D bosonic topological phases protected by the inversion symmetry.

%So the 1D bosonic topological phases protected by the inversion symmetry are directly realized by loading the interacting bosons into the corresponding optical %superlattices.

%%%%%%%%%%%%%%%%%%%
\textit{Summary-} In summary, we have identified various SPT phases in the 1D models with periodically modulated hopping amplitudes or on-site potentials. The topological phase is characterized by a nontrivial Berry phase or a pair of degenerate in-gap boundary states. The symmetries protecting the topological phases are explicitly analyzed and the connection between the two kinds of modulations is discussed. We also find that the 1D bosonic topological phase protected by the inversion symmetry can be realized directly by loading the interacting bosons into the same lattices. The identified SPT phases are possible to be observed in the superlattice systems which are realizable in current cold atomic experiments or photonic crystal setups.
%%%%%%%%%%%%%%%%%%%

%%%%%%%%%%%%%%%%%%%
\textit{Acknowledgments-}
%%%%%%%%%%%%%%%%%%%
H.G. is supported by NSFC under Grants No.11274032 and No. 11104189,
FOK YING TUNG EDUCATION FOUNDATION, Program for NCET. S. C. is supported by NSFC under Grants No. 11374354, No. 11174360, and No. 11121063, and by the Strategic Priority Research Program of the Chinese Academy of Sciences under Grant No. XDB07000000.

\appendix

\clearpage

\section*{Supplemental Material}
In the supplemental materia, in order to see clearly the localization properties of the in-gap boundary states, we display their density distributions for systems with different $\lambda$. The distributions are calculated via $n_i=\langle \psi_{bs}|\hat{n}_i|\psi_{bs}\rangle$ with $\hat{n}_i=\hat{c}_i^{\dagger}\hat{c}_i$ the number operator and $\psi_{bs}$ the eigenvector of the boundary state. Since the two degenerate boundary states are related by the symmetry protecting the topological phase, only one of them is shown. The distribution of the boundary state at $1/3$ filling and $\phi=0$ corresponding to the one in Fig.\ref{fig1}(e) with $\lambda=0.6$ is shown in Sfig.1. Its distribution localizes near the boundaries with a tail extending into the bulk, whose extent depends on the bulk gap. As a comparison, we also study the case of $\lambda=0.3$ with a smaller gap. Since the bulk gap is smaller, the boundary state becomes less localized. The situations are the same for the boundary states of the $T=4$ case at $1/4$ and $1/2$ fillings. The case at $1/4$ filling with the same parameters as those of $T=3$ case is shown in Sfig.1 (b). Compared to the $T=3$ case at $1/3$ filling, since the bulk gaps only have small difference, the localizations are similar though the details differ. At $1/2$ filling and with the parameters used in Fig.\ref{fig1}(f), the in-gap states also distribute near the boundaries and more components extend into the bulk when the bulk gap becomes smaller [see Sfig.1 (c)]. The results for the model with periodically modulated potentials are similar and are not shown here and thereafter.

\begin{figure}[htbp]
\centering \includegraphics[width=8.5cm]{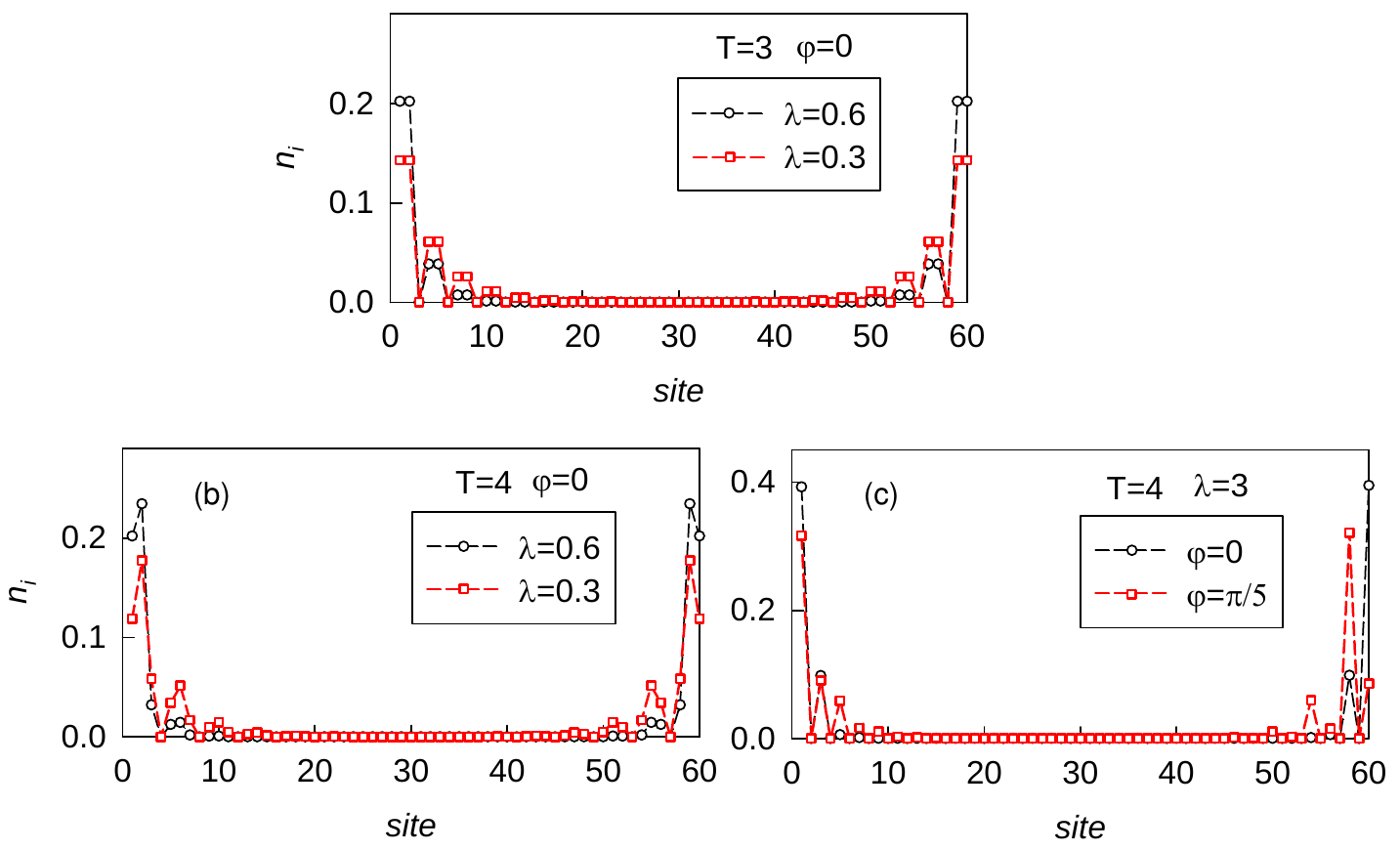} \caption{Sfig.1: (Color online) The distribution of the boundary state: (a) $T=3$ at $1/3$ filling; (b) $T=4$ at $1/4$ filling; (c) $T=4$ at $1/2$ filling. The size of the open chain is $L=60$.}
\label{sfig1}
\end{figure}

In the main text the total size $L$ is taken as $L=NT$ with $N$ the number of unit cells. One can also take the lattice size as $L=NT+q$ with $q=1,...,T-1$. On an open chain with the size $L=NT+q$, there appear a pair of degenerate boundary states at $\varphi=-\frac{q\pi}{T}$ for $1/T$ and $(T-1)/T$ fillings, when the hopping amplitudes $t_{j,j+1}$ are inversion symmetric and the 1D topological phase protected by the inversion symmetry is realized. Though the system is lack of the periodicity, it can be viewed as a supercell of an infinite system and the Berry phase is defined using the twisted boundary phase: $\gamma=\oint i\langle \psi_{\theta}|\frac{d}{d\theta}|\psi_{\theta}\rangle$, where $\theta$ is the twisted boundary phase which takes values from $0$ to $2\pi$ and $\psi_{\theta}$ is the corresponding ground-state wave function.

In Sfig.2, we show the results of the $T=3$ case with the parameters used in Fig.\ref{fig1} (e) except that the length is $L=61$ $(q=1)$.
The degenerate boundary states appear at $\varphi=-\pi/3$, at which the hopping amplitudes are inversion symmetric and the Berry phase defined using the twisted boundary phase has the nontrivial value $\pi$. The in-gap state, which is due to the nontrivial topological property, is localized near the boundaries. The results of $q=2$ are similar.

We also show the results of the $T=4$ case with the parameters used in Fig.\ref{fig1} (f) except the length of the system. We consider two typical lengthes, i.e., $L=61$ $(q=1)$ and $L=62$ $(q=2)$. The results at $1/4$ and $3/4$ fillings are similar with those of the $T=3$ case. The chiral symmetry is broken (or not) when the length is odd (or even). The topological phases protected by the chiral symmetry at half filling remain on chains with even number of sites [see Sfig.3(b)] and the in-gap states distribute near the boundaries [see Sfig.3(d)]. It is noted in Sfig.3(a) that there appears a in-gap state on the $L=61$ chain. However it is not degenerate due to the breaking of the chiral symmetry for the odd lattice.
%In fact it exists even when the twisted phase is added.

\begin{figure}[htbp]
\centering \includegraphics[width=8.5cm]{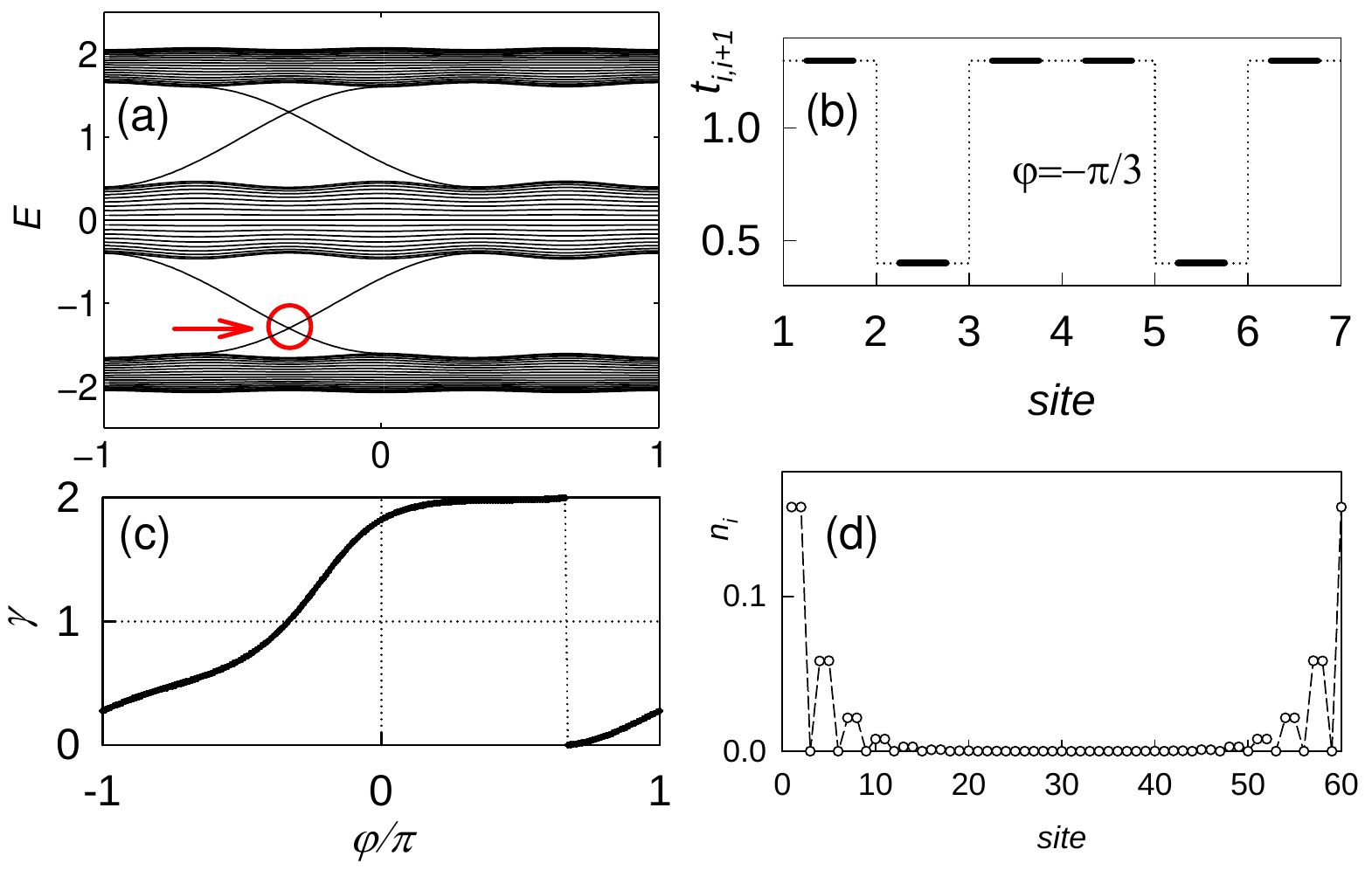} \caption{Sfig.2: (Color online) (a) The energy spectrum on an open chain with the period $T=3$ and the length $L=61$. (b) The hopping amplitudes at inversion symmetric point $\varphi=-\pi/3$. (c) The Berry phase corresponding to the above figure at $1/3$ filling. (d) The distribution of the boundary state denoted by the red arrow in (a). The parameter $\lambda=0.6$, which is the same as the one used in Fig.1(e).}
\label{sfig2}
\end{figure}

\begin{figure}[htbp]
\centering \includegraphics[width=8.5cm]{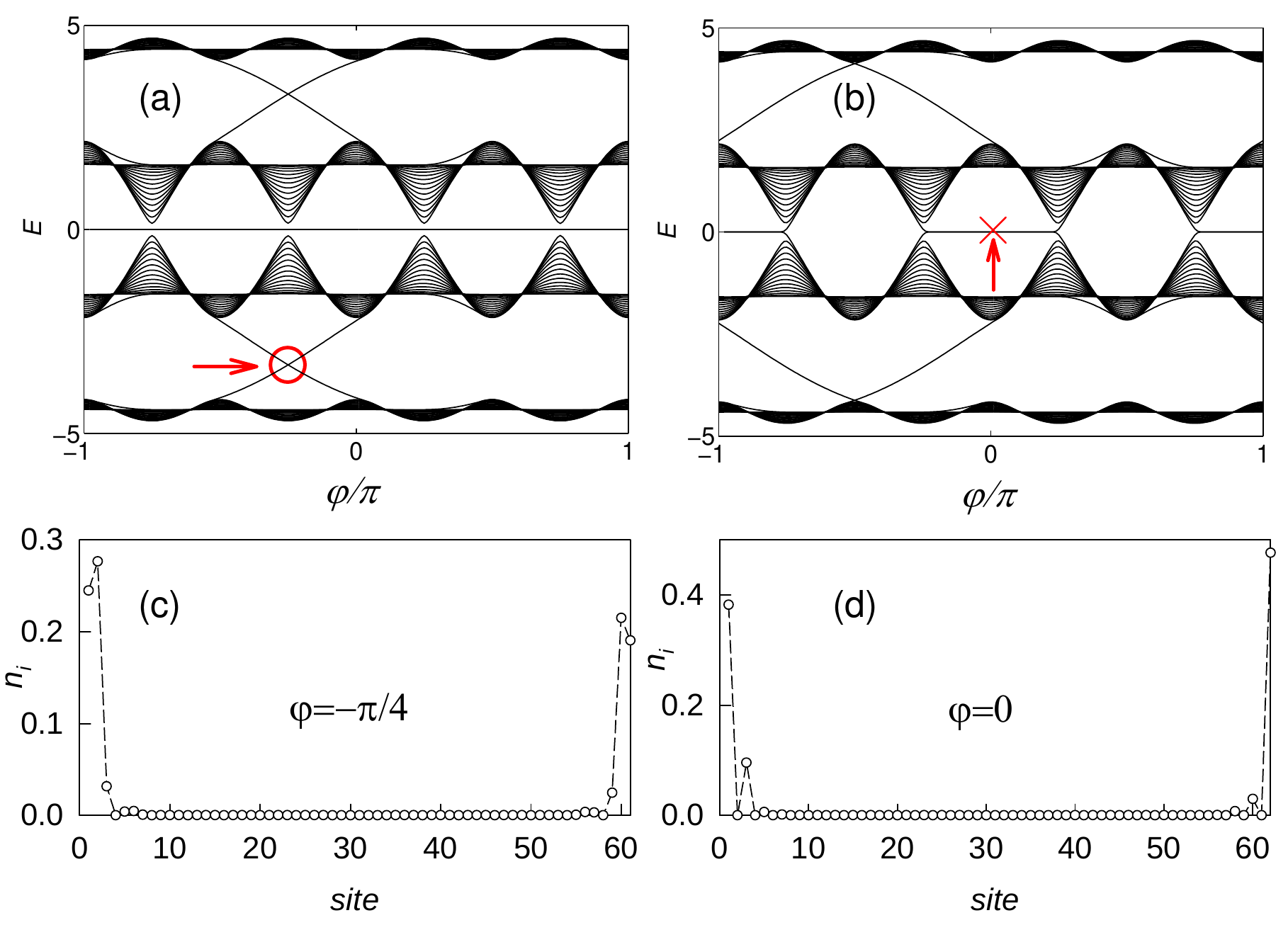} \caption{Sfig.3: (Color online) The energy spectrum on an open chain with the period $T=4$ and the length (a) $L=61$; (b) $L=62$. (c) and (d): the distribution of the boundary state denoted by the red arrow in the corresponding above figure. The parameter $\lambda=3$, which is the same as the one used in Fig.1(f).}
\label{sfig3}
\end{figure}

In Eq.(\ref{eq2}) of the main text, the Hamiltonian in the momentum space of the $T=4$ system is given. It is derived in the basis $(\hat{c}_{1,k},\hat{c}_{2,k},\hat{c}_{3,k},\hat{c}_{4,k})^{T}$ and is the following $4\times 4$ matrix:
\begin{eqnarray*}\label{mrt}
\hat{H}_1(k)=\left(
  \begin{array}{cccc}
    0 & t_1 & 0 & t_4 e^{-i k} \\
    t_1 & 0 & t_2 & 0 \\
    0 & t_2 & 0 & t_3 \\
    t_4 e^{i k} & 0 & t_3 & 0 \\
  \end{array}
\right)
\end{eqnarray*}
In terms of the Dirac matrices, which are the Kronecker product of two Pauli matrices, it can be wrote compactly as the one in Eq.(\ref{eq2}). The derivation of Eq.(\ref{eq4}) is similar.

\begin{thebibliography}{10}
\bibitem{notep} The results with periodical modulations of on-site potentials are similar.
\bibitem{note} The energy gap is taken as the minimum of those under different twisted boundary phases, which greatly reduces the finite-size effect.
\end{thebibliography}
\end{document}